\documentclass[prl,twocolumn,showpacs,superscriptaddress]{revtex4}

\usepackage{amsmath}
\usepackage{amssymb}
\usepackage{graphicx}

\begin{document}

\title{Momentum-Resolved Landau-Level Spectroscopy of Dirac Surface State in Bi$_2$Se$_3$}

\author{T. Hanaguri}
\affiliation{RIKEN Advanced Science Institute, Wako, Saitama 351-0198, Japan}

\author{K. Igarashi}
\affiliation{Materials and Structures Laboratory, Tokyo Institute of Technology, Yokohama, Kanagawa 226-8503, Japan}

\author{M. Kawamura}
\affiliation{RIKEN Advanced Science Institute, Wako, Saitama 351-0198, Japan}

\author{H. Takagi}
\affiliation{RIKEN Advanced Science Institute, Wako, Saitama 351-0198, Japan}
\affiliation{Department of Advanced Materials, University of Tokyo, Kashiwa, Chiba 277-8561, Japan}

\author{T. Sasagawa}
\affiliation{Materials and Structures Laboratory, Tokyo Institute of Technology, Yokohama, Kanagawa 226-8503, Japan}

\date{\today}

\begin{abstract}
We investigate Dirac fermions on the surface of the topological insulator Bi$_2$Se$_3$ using scanning tunneling spectroscopy.
Landau levels (LLs) are observed in the tunneling spectra in a magnetic field.
In contrast to LLs of conventional electrons, a field independent LL appears at the Dirac point, which is a hallmark of Dirac fermions.
A scaling analysis of LLs based on the Bohr-Sommerfeld quantization condition allowed us to determine the dispersion of the surface band.
Near the Fermi energy, fine peaks mixed with LLs appear in the spectra, which may be responsible for the anomalous magneto-fingerprint effect [J. G. Checkelsky {\it et al.}, Phys. Rev. Lett. {\bf 103}, 246601 (2009)].
\end{abstract}


\maketitle

The recent development of spintronics has drawn much attention to spin-orbit coupling in metals and semiconductors.
One of the most striking discoveries which has emerged from research into spin-orbit coupling is a new state of matter known as a topological insulator (TI)~\cite{TI_Reviews}.
Combined with time-reversal symmetry, spin-orbit coupling generates a robust conducting state at the boundary, whereas there is an energy gap in the bulk.

The band dispersion of the boundary state is linear in momentum $k$, forming a massless Dirac cone on the surface of three-dimensional TI.
Dirac cones are also realized in other materials including graphene~\cite{CastroNeto2009RMP}, organic conductors~\cite{Tajima2007EPL} and $d$-wave superconductors~\cite{Balatsky2006RMP} but the Dirac surface state of TI has some unique features~\cite{TI_Theories}.
First, the Dirac cone of TI is centered at the time-reversal invariant momentum (TRIM), which guarantees the robustness of the Dirac spectrum against any time-reversal invariant perturbations.
Second, in TI, a single Dirac cone can be formed in the surface Brillouin zone.
Finally, because of the strong spin-orbit coupling, spins are locked with momenta, giving rise to helical Dirac fermions without spin degeneracy.
In contrast to these features, Dirac cones in graphene, for example, are not at the TRIM, always come in pairs, and have a spin degeneracy.
A Dirac cone at the TRIM was observed in some TIs including Bi-Sb~\cite{Hsieh2008Nature}, Bi$_2$Te$_3$~\cite{Chen2009Science} and Bi$_2$Se$_3$~\cite{Xia2009NatPhys} by angle-resolved photoemission spectroscopy (ARPES).
Helical spin structure in $k$ space was recently verified by spin-resolved ARPES~\cite{Hsieh2009Nature,Nishide2009PRB}.

To capture the physics of Dirac fermions in TI further, we study the response of Dirac fermions to defects and magnetic field using low-temperature scanning tunneling microscopy/spectroscopy (STM/STS).
We measure Bi$_2$Se$_3$ which is the simplest TI with only one surface state forming a single isotropic Dirac cone at the center of the Brillouin zone~\cite{Zhang2009NatPhys}.
Naturally grown Bi$_2$Se$_3$ crystals are doped with electrons due to unavoidable non-stoichiometry, predominantly  Se vacancies.
Typically, the Dirac point, the tip of the cone, is located a few hundred meV below $E_F$ and the bulk conduction band crosses $E_F$~\cite{Xia2009NatPhys,Analytis2010condmat,Park2010PRB}.
Such a situation makes it difficult to observe the signals from Dirac fermions by transport experiments.
STM/STS has an advantage that a wide energy range including the energy of Dirac point $E_{DP}$ can be explored.

Experiments have been conducted on (111) surfaces of Bi$_2$Se$_3$ single crystals prepared by a melt-growth technique.
Two samples, Sample~1 and Sample~2, from different growth batches were used.
Resistivity and Seebeck-coefficient measurements suggest that Sample~2 contains less Se vacancies than Sample~1.
These samples were cleaved {\it in situ} at 77~K under ultrahigh vacuum and were immediately transferred to the STM unit kept at cryogenic temperature below 10~K.
Differential conductance spectrum $g({\bf r},E)$, which reflects the local density of states (LDOS) at location {\bf r} and energy $E$, was measured by the standard lock-in technique.
All the STM/STS data were collected at 1.5~K under magnetic fields perpendicular to the surfaces.

We show an STM topograph of Sample~1 in Fig.~1(a).
Triangular-shaped defects with different intensities are superposed on a weak atomic-lattice corrugation with an amplitude of several picometers and a periodicity of 0.4~nm.
These defects may represent Se vacancies and/or Bi defects at the Se sites at different layers.
In Fig.~1(b) we plot a typical LDOS spectrum.
The LDOS spectrum exhibits a V-shape with a minimum at -290~mV, representing $E_{DP}$.
Below about -400~mV and above $E_F$, the LDOS increases rapidly, likely associated with contributions from the bulk bands.
There is a finite offset even at $E_{DP}$ whose origin is unclear at present.
All of these features agree well with previous STM/STS results~\cite{Urazhdin_STM}.
\begin{figure}
\includegraphics[width=85mm]{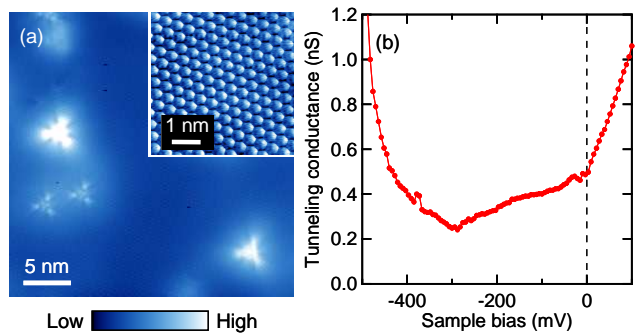}
\caption{(color online).
(a)
STM topograph of cleaved (111) surface of Bi$_2$Se$_3$ showing triangular-shaped defects.
The set-up conditions for imaging were sample-bias voltage $V_s=-100$~mV and tunneling current $I_t=0.1$~nA.
Similar image was observed at $V_s=+100$~mV, suggesting that topographic effect plays major role for the defect image.
The inset shows a magnified topograph showing the atomic resolution (taken at $V_s=+50$~mV, $I_t=0.5$~nA).
(b)
Tunneling spectrum taken at the center of Fig.~1(a) measured with a bias modulation amplitude $V_{\rm mod}=3.5$~mV$_{\rm rms}$.
Junction resistance was set to 1~G$\Omega$ at $V_s=+100$~mV.
The V-shaped spectrum is consistent with the Dirac dispersion with $E_{DP}=-290$~mV.
}
\end{figure}

Let us first focus on the states near the defects.
Defects may scatter quasi-particles and generate quasi-particle interference (QPI) modulations in the $g({\bf r},E)$ maps.
QPI modulations have been observed in various materials and have been used to explore $k$ space by STM/STS~\cite{QPI_Experiments}.
The scattering vectors which contribute to QPI connect the states in $k$ space with the same energy and the same spin direction.
Although the spin selectivity does not appear in a spin degenerate system, it plays a crucial role in TIs where the spin degeneracy is lifted~\cite{Roushan2009Nature,Zhang2009PRL,Alpichshev2010PRL,Gomes2009condmat}.
In the case of Bi$_2$Se$_3$, the spin selectivity may diminish the QPI, because the backscattering channel always connects the states with the opposite spins in a single isotropic Dirac cone.
In order to confirm this, we map normalized conductance $g({\bf r},E)/<g({\bf r},E)>$ in Figs.~2(a)-(c).
Here $<...>$ denotes the spatial average and these maps were taken in the same field of view of Fig.~1(a).
According to ARPES~\cite{Xia2009NatPhys}, the Fermi momentum of the surface state in Bi$_2$Se$_3$ with similar $E_{DP}$ is about 1~nm$^{-1}$.
In the absence of the spin selectivity, this number would yield a QPI modulation with a periodicity of 3~nm at $E_F$ which becomes shorter with increasing $E$.
In reality, however, $g({\bf r},E)/<g({\bf r},E)>$ is rather uniform except for weak features at the defects and no appreciable QPI modulations are detected.
We quantify the homogeneity by calculating histograms of $g({\bf r},E)/<g({\bf r},E)>$.
As shown in Fig.~2(d), the histograms exhibit very sharp peak at unity with a full width at half maximum less than 10~\%.
Such a homogeneous LDOS in Bi$_2$Se$_3$ is in sharp contrast to the cases of other TIs, Bi-Sb with multiple surface states~\cite{Roushan2009Nature,Gomes2009condmat} and Bi$_2$Te$_3$ with the warped Fermi surface~\cite{Zhang2009PRL,Alpichshev2010PRL}, and indicates that QPI is absent in a single, isotropic and helical Dirac cone.
\begin{figure}
\includegraphics[width=85mm]{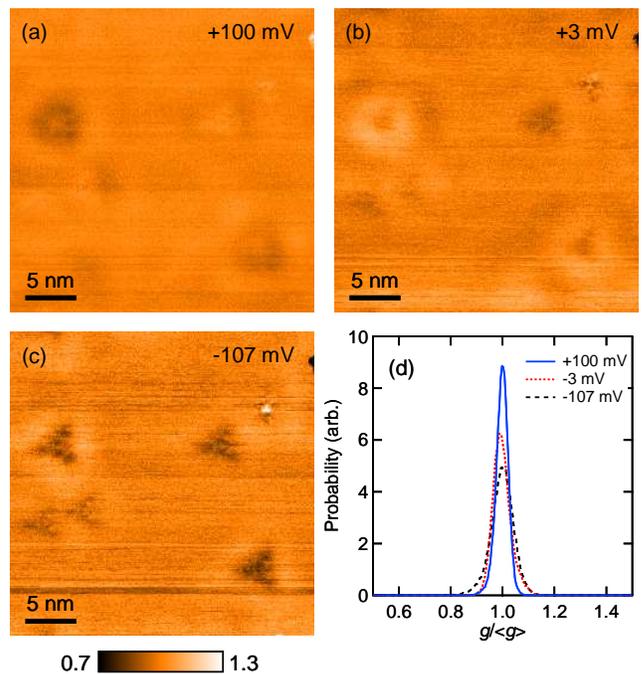}
\caption{(color online).
(a)-(c)
Normalized conductance maps $g({\bf r},E)/<g({\bf r},E)>$ imaged in the same field of view of Fig.~1(a).
Set up conditions for the measurement are the same as those of Fig.~1(b).
Weak features are observed at the defects but clear QPI modulations are not detected.
(d)
Histograms of $g({\bf r},E)/<g({\bf r},E)>$ shown in Figs.~2(a)-(c).
Peaks have Gaussian shape with a full width at half maximum less than 10~\% of mean value.
Such a uniform $g({\bf r},E)/<g({\bf r},E)>$ suggests that the QPI is suppressed in Bi$_2$Se$_3$.
}
\end{figure}

Given the fact that QPI is unavailable in Bi$_2$Se$_3$, we develop an alternative way to study the $k$-space electronic states using STM/STS.
We pay attention to the Landau levels (LLs) which appear in the LDOS spectrum in a magnetic field.
The LL in TIs is very important because the anomalous quantum nature of Dirac fermions becomes obvious in a magnetic field~\cite{CastroNeto2009RMP,Li2009PRL,Zheng2002PRB}.
The energy $E_n$ of $n$-th LL of Dirac fermions is expressed as
\begin{equation}
E_n=E_{DP}+{\rm sgn}(n)v\sqrt{2e\hbar|n|B},
\label{E_n}
\end{equation}

\noindent
where $n$ ($=0,\pm1,\pm2,...$) is the LL index, $v$ is the velocity of electron, $e$ is the unit charge, $\hbar$ is the Plank constant and $B$ is the magnetic field~\cite{Zheng2002PRB}.
An important feature is that there is a $B$-independent $n=0$ LL at $E_{DP}$ which is absent in a conventional massive electron system.
In addition, $E_n$ is not linear in $nB$ but is proportional to $\sqrt{|n|B}$ so long as $v$ is constant over the energy range of interest.
In graphene, these features are observed by STM/STS very clearly~\cite{Li2009PRL}.
In Bi$_2$Se$_3$, however, constant $v$ approximation breaks down as ARPES observes significant deviation from the linear $E$-$k$ dispersion relation~\cite{Xia2009NatPhys}.

Here we demonstrate a way to derive the $E$-$k$ dispersion from the LL spectroscopy.
Since the Landau orbit is quantized in $k$ space, we can pick out specific momentum.
According to the Bohr-Sommerfeld quantization condition, the area $S_n$ of the $n$-th Landau orbit in $k$ space is given by $S_n=(n+\gamma)2\pi eB/\hbar$, where $\gamma$ is the phase factor which is 1/2 for conventional free electrons.
In Dirac fermions, the quantum-mechanical Berry-phase effect eliminates $\gamma$, giving rise to the $n=0$ LL at $E_{DP}$~\cite{Zheng2002PRB}.
Since the Dirac cone is isotropic in Bi$_2$Se$_3$, $S_n$ can simply be expressed by the single momentum $k_n$ as $S_n=\pi k_n^2$, and we get
\begin{equation}
k_n=\sqrt\frac{2e|n|B}{\hbar}.
\label{k_n}
\end{equation}

\noindent
Thus, once we specify $n$ and $B$, a set of $E_n$ and $k_n$ can be obtained.
Since $\sqrt{|n|B}$ represents $k_n$, $E_n$ should be scaled by $\sqrt{|n|B}$ and the scaling function is nothing but the $E$-$k$ dispersion relation.

Figure~3(a) shows tunneling spectra of Sample~2 taken at different magnetic fields.
$E_{DP}$ or the bottom of the V-shaped spectrum at $B=0$~T is shifted closer to $E_F$ as compared to Sample~1 [Fig.~1(b)], which is consistent with the reduction of Se vacancies in Sample~2.
Above $\sim$4~T, sequential LDOS peaks are clearly observed, which we assign to LLs.
More than 20 levels can be identified at higher fields.
A series of LLs were also observed in Sample~1 despite the LDOS peaks being slightly blurred.
There appears a peak at $E_{DP}$ of which position does not depend on $B$.
This corresponds to the $n=0$ LL, an unambiguous indication that the observed LLs are indeed associated with the Dirac surface state.

The peak positions, $E_n$, were determined by fitting the spectrum with multiple Lorentzian functions.
As shown in the inset of Fig.~3(b), $E_n$ is sub-linear in $n$, which is another important consequence of Dirac fermions.
In order to test the expected scaling behavior, we plot $E_n$ against $\sqrt{|n|B}$ in Fig.~3(b).
The scaling relation holds quite well in both samples.
The scaling parameter $\sqrt{|n|B}$ can be converted into the momentum by Eq.~\ref{k_n} as shown in the bottom axis of Fig.~3(b).
It is obvious that the dispersions of Sample~1 and Sample~2 are essentially the same despite the difference in $E_{DP}$, implying a rigid-band shift associated with electron doping.
The obtained $E$-$k$ dispersion is a convex function which coincides in detail with the ARPES result~\cite{Xia2009NatPhys}, demonstrating the validity of the $k$-resolved LL spectroscopy.
\begin{figure}
\includegraphics[width=85mm]{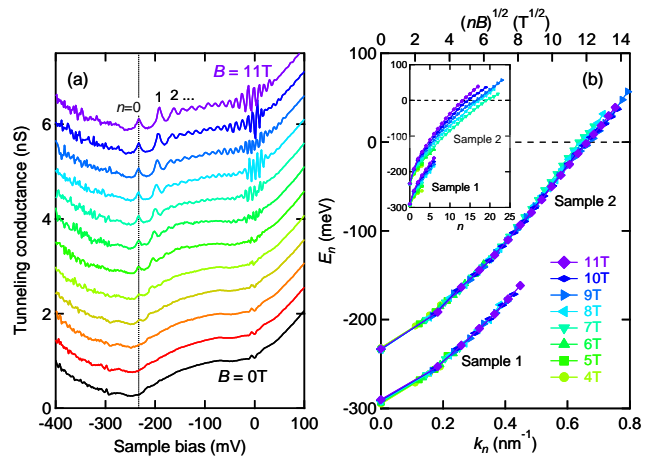}
\caption{(color online).
(a)
Tunneling spectra showing the series of peaks associated with the LL formation.
Data were collected in a magnetic field perpendicular to the surface from 0~T up to 11~T with 1~T interval.
$V_s=+100$~mV, $I_t=0.2$~nA and $V_{\rm mod}=1.8$~mV$_{\rm rms}$.
Spectra under the fields are shifted vertically for clarity.
Note that there is a field-independent $n=0$ LL at $E_{DP}$ indicated by a dotted line.
(b)
Energy of LLs $E_n$ scaled by the square-roots of field $B$ and LL index $n$.
The scaling parameter $\sqrt{|n|B}$ (top axis) can be converted into the momentum $k$ (bottom axis) and the scaling function represents band dispersion.
See text for details.
The inset shows the $n$ dependence of $E_n$ before the scaling.
}
\end{figure}

Although information on the in-plane anisotropy is not available, this technique based on LL spectroscopy can be useful not only for TI but also for a variety of systems with a small Fermi pocket.
In order to resolve the small Fermi pocket, ARPES in fact must be performed with very high energy and momentum resolutions.
The QPI technique can be employed instead, if there is no spin degeneracy.
However, a very wide field of view should be scanned to resolve small momenta.
The LL spectroscopy is free from these constrains.

By inspecting the details of LDOS spectra, we notice some unusual aspects which cannot be understood within the simple Dirac-fermion model.
First, LLs with $n<0$ are absent.
Slower $v$ below $E_{DP}$~\cite{Xia2009NatPhys} may partly account for the missing LLs but it would be necessary to assume excess quasi-particle damping to diminish the LL peaks completely.
Coupling to the bulk valence band located just below $E_{DP}$~\cite{Xia2009NatPhys} is one of the possible sources of damping.
It may be informative here to point out that, near $E_F$ where the bulk {\it conduction} band overlaps with the surface state ~\cite{Xia2009NatPhys,Analytis2010condmat,Park2010PRB}, LLs are clearly observed and even enhanced as shown in Fig.~3(a).
Such contrasting behaviors may indicate that the coupling between the Dirac surface state and the bulk bands is qualitatively different between the valence and the conduction bands.

The enhancement of the LL peaks near $E_F$ suggests that the quasi-particle lifetime becomes longer on approaching $E_F$.
In order to confirm this, we extract the peak width of each LL by a Lorentzian fitting.
As shown in Fig.~4(a), the peak width shows a minimum at $E_F$.
This is direct evidence that the quasi-particle lifetime increases near $E_F$.
Analogous observation was also reported in graphene, and was attributed to electron-electron interaction~\cite{Li2009PRL}.
In addition to the minimum at $E_F$, the peak width has a broad maximum at about -150~meV.
Such a complicated energy dependence of the peak width may indicate that there is more than one quasi-particle relaxation process, as also suggested by the recent ARPES experiment~\cite{Park2010PRB}.

Yet another anomaly was found in the close vicinity of $E_F$.
As shown in Fig.~4(b), high-resolution LDOS spectra exhibit fine peaks in the energy range between $\sim\pm$20~mV.
The fine peaks systematically shift to higher energies with increasing magnetic field in the same manner as LLs of the surface state.
This suggest that the Dirac fermions at the surface are responsible for the formation of fine peaks.
We note that linear superposition of LLs of the bulk conduction band cannot account for the observation because the Fermi surface of the bulk band is smaller than that of the surface state~\cite{Xia2009NatPhys,Hsieh2009Nature}, being incompatible with the finer structure.
In the energy range where the fine peaks appear, the amplitudes of the peaks become large and resemble an envelope function [Fig.~4(c)].
It is clear that the energy range between $\sim\pm$20~mV differs from the other energy regions and the electronic states there cannot be described by the LLs alone.

Although the exact nature of electronic states near $E_F$ is unclear at present, the observed fine peaks should manifest itself in the magneto-resistance because LDOS at $E_F$ is directly related to the transport properties.
Indeed, an anomalously large amplitude magneto-fingerprint effect has been observed in Bi$_2$Se$_3$ and is interpreted as a result of non-trivial coupling between the surface and the bulk~\cite{Checkelsky2009PRL}.
Further experimental and theoretical efforts are necessary to clarify the origin of the fine peaks and their relation to the magneto-fingerprint effect.
\begin{figure}
\includegraphics[width=85mm]{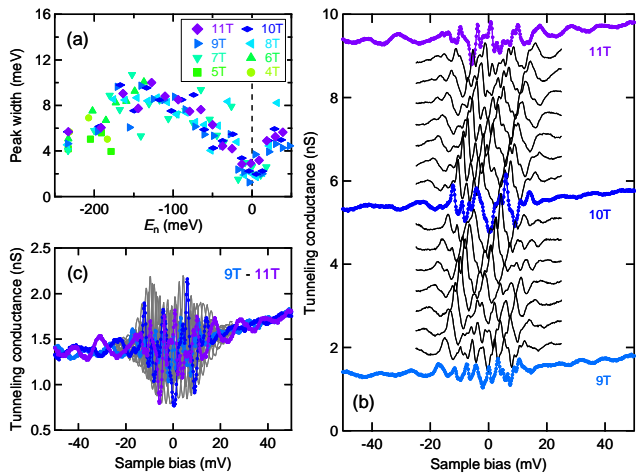}
\caption{(color online).
(a)
Full width at half maximum of the fitted Lorentzian function for each LL plotted as a function of $E_n$.
(b)
High-resolution tunneling spectra near $E_F$ showing the extra fine structures mixed with LLs.
$V_s=+100$~mV, $I_t=0.2$~nA and $V_{\rm mod}=0.3$~mV$_{\rm rms}$.
Measurements were made on a different cleave of Sample 2 in a magnetic field from 9~T to 11~T with 0.125~T interval.
Spectra are shifted vertically for clarity.
(c)
Spectra shown in Fig.~4(b) are plotted without shift.
}
\end{figure}

In summary, we have studied a prototypical TI, Bi$_2$Se$_3$, using STM/STS in a magnetic field and have revealed several important aspects of the Dirac surface state.
The QPI modulations around defects are not detected, which we argue is the consequence of suppressed backscattering associated with the helical spin structure in $k$ space.
We have succeeded in observing a series of LLs which possess the unique characteristics of Dirac fermions including a $B$-independent $n=0$ LL at $E_{DP}$.
The band dispersion of the surface state has been determined from the energy of LLs using an analysis based on the Bohr-Sommerfeld quantization condition.
In addition to the LLs, LDOS spectra under magnetic fields exhibit unusual features including the suppression of quasi-particle damping and extra fine structures near $E_F$.
We anticipate that these observations provide a spectroscopic basis for the understanding of Dirac fermions in TI.

The authors thank A. V. Balatsky, A. F. Bangura, A. Furusaki, S. Murakami, D. Rees, T. Sasaki and S. -C. Zhang for valuable discussions and comments.
This work has been supported by Grant-in-Aid for Scientific Research from the Ministry of Education, Culture, Sports, Science and Technology of Japan.

{\it Note added.-} During the completion of the manuscript, we became aware that Cheng {\it et al.} also succeeded in observing the LLs in Bi$_2$Se$_3$ thin films~\cite{Cheng2010condmat}.

\end{document}